\documentclass{moriond}

\def\be{\begin{equation}}
\def\ee{\end{equation}}
\def\bea{\begin{eqnarray}}
\def\eea{\end{eqnarray}}

\usepackage{amsmath,amssymb}
\usepackage{hepnames}
\usepackage{graphicx}
\usepackage{cite}
\usepackage{xspace}
\usepackage{booktabs}
\usepackage{microtype}

\newcommand{\fbinv}{\ensuremath{\,\text{fb}^{-1}}\xspace}
\newcommand{\fb}{\ensuremath{\,\text{fb}}\xspace}
\newcommand{\pb}{\ensuremath{\,\text{pb}}\xspace}
\newcommand{\tev}{\ensuremath{\,\text{TeV}}\xspace}
\newcommand{\gev}{\ensuremath{\,\text{GeV}}\xspace}
\newcommand{\mjj}{\ensuremath{m_{jj}}\xspace}
\newcommand{\detajj}{\ensuremath{\Delta\eta_{jj}}\xspace}
\newcommand{\pTZZ}{\ensuremath{p_{\mathrm{T}}^{ZZ}}\xspace}
\newcommand{\muEWK}{\ensuremath{\mu_{\mathrm{EWK}}}\xspace}
\newcommand{\muQCD}{\ensuremath{\mu_{\mathrm{QCD}}}\xspace}
\newcommand{\ONN}{\ensuremath{O_{\mathrm{NN}}}\xspace}
\newcommand{\sigmaEW}{\ensuremath{\sigma_{\mathrm{EW}}}\xspace}

\newcommand{\Photo}{}

\begin{document}
\vspace*{4cm}
\title{Multiboson and VBS measurements in ATLAS and CMS}
\author{ S. Folgueras, on behalf of the CMS and ATLAS Collaborations\footnote{Copyright 2026 CERN for the benefit of the ATLAS and CMS Collaborations. Reproduction of this article or parts of it is allowed as specified in the CC-BY-4.0 license}}

\address{University of Oviedo - ICTEA \\
Oviedo, Spain}

\maketitle\abstracts{A review of recent multiboson and vector boson scattering (VBS) measurements from the ATLAS and CMS Collaborations at the LHC is presented. Results are reported from precision diboson cross-section measurements, novel CP-sensitive and polarisation observables in $W\gamma$ production, VBS observations in semileptonic and fully leptonic final states including the first measurements at $\sqrt{s}=13.6\tev$, and observations of triboson processes. These results constitute a comprehensive test of the electroweak gauge sector of the Standard Model, and provide stringent constraints on anomalous gauge couplings in the effective field theory framework.}

\section{Introduction}

Multiboson production processes at the LHC provide a unique window into the non-Abelian gauge structure of the Standard Model (SM). The measurement of diboson, triboson, and vector boson scattering processes serves three interconnected goals: testing triple and quartic gauge couplings (TGC/QGC), probing the interplay between the gauge sector and the Higgs boson, and searching for new physics through deviations from SM predictions, parametrised in an effective field theory (EFT) framework using dimension-6 and dimension-8 operators.

Vector boson scattering processes --- characterised by two vector bosons produced in association with two forward jets, with large dijet invariant mass \mjj and large rapidity separation \detajj --- represent a particularly privileged topology: the cancellation between gauge and Higgs exchange diagrams that ensures perturbative unitarity at high energy is directly accessible experimentally. Following the Run~2 programme, VBS has been established in essentially all major channels, and Run~3 is now opening the door to precision cross-section measurements and enhanced EFT reach.

In this review, results from ATLAS~\cite{ATLAS:2008xda} and CMS~\cite{CMS:2008xjf} using Run~2 data collected at $\sqrt{s}=13\tev$ (up to $140\fbinv$) and first Run~3~\cite{CMS:2023gfb} results at $\sqrt{s}=13.6\tev$ (up to $280\fbinv$) are presented. All measured cross sections and signal strengths are found to be consistent with the SM expectation within uncertainties.

\section{Precision Diboson Measurements}
Diboson processes constitute the baseline of the electroweak precision programme. Their production cross sections are now known at NNLO QCD accuracy, and the growing datasets enable differential measurements that are sensitive to the details of higher-order EW corrections.

\subsection{Differential \texorpdfstring{$ZZ$}{ZZ} production at 13 TeV (ATLAS)}

ATLAS has measured $ZZ\to\ell\ell\nu\nu$ production at $\sqrt{s}=13\tev$ using the full Run~2 dataset of $140\fbinv$~\cite{ATLAS:ZZ2025}. The analysis covers both inclusive $ZZ$ production and $ZZ$ production in association with two jets ($ZZjj$), in a phase space dominated by the VBS topology. The measured fiducial cross sections are:
\begin{align}
\sigma_{\rm fid}(ZZ \to \ell\ell\nu\nu) &= 21.0 \pm 1.0\fb, \\
\sigma_{\rm fid}(ZZjj \to \ell\ell\nu\nu jj) &= 0.96\,^{+0.18}_{-0.16}\fb.
\end{align}
Extrapolated to the full phase space with $66 < m_{\ell\ell} < 116\gev$, the inclusive cross section is $\sigma(pp\to ZZ) = 15.38 \pm 0.81\pb$, in good agreement with the NNLO QCD prediction. Differential distributions in eight kinematic observables, including \pTZZ up to $1.5\tev$ in the $ZZjj$ channel, are in agreement with NLO predictions from MATRIX and Sherpa+MG5. The high-\pTZZ tails are exploited to set constraints on neutral anomalous TGC operators via both the effective vertex approach and the SMEFT framework.

\subsection{\texorpdfstring{$W\gamma$}{Wgamma} production: polarisation and CP-sensitive observables (ATLAS)}

\begin{figure}
    \centering
    \includegraphics[width=0.6\linewidth]{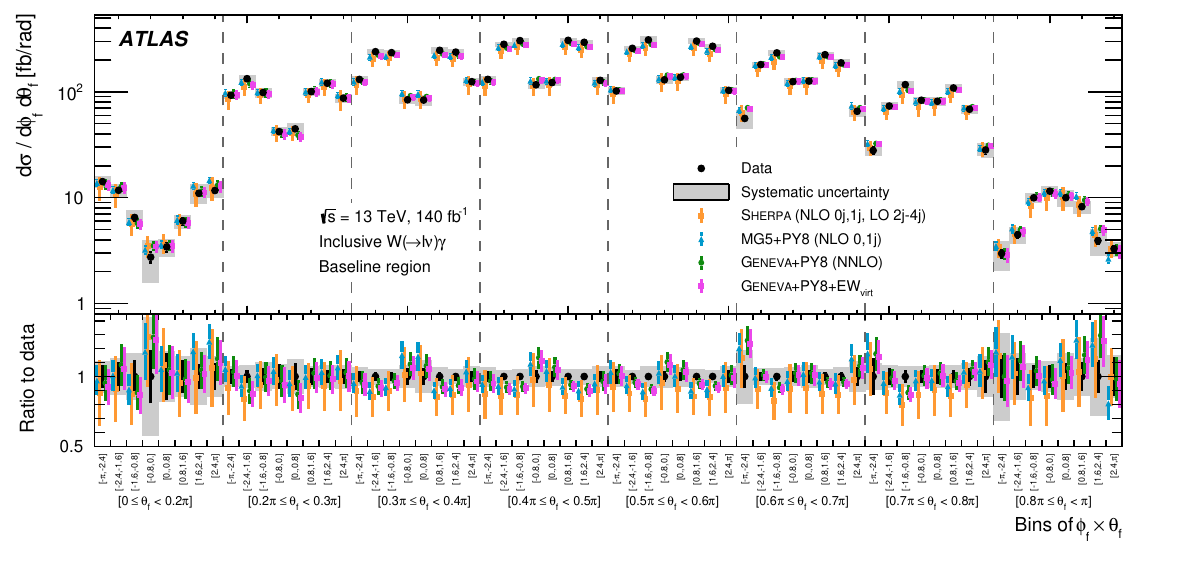}
    \includegraphics[width=0.3\linewidth]{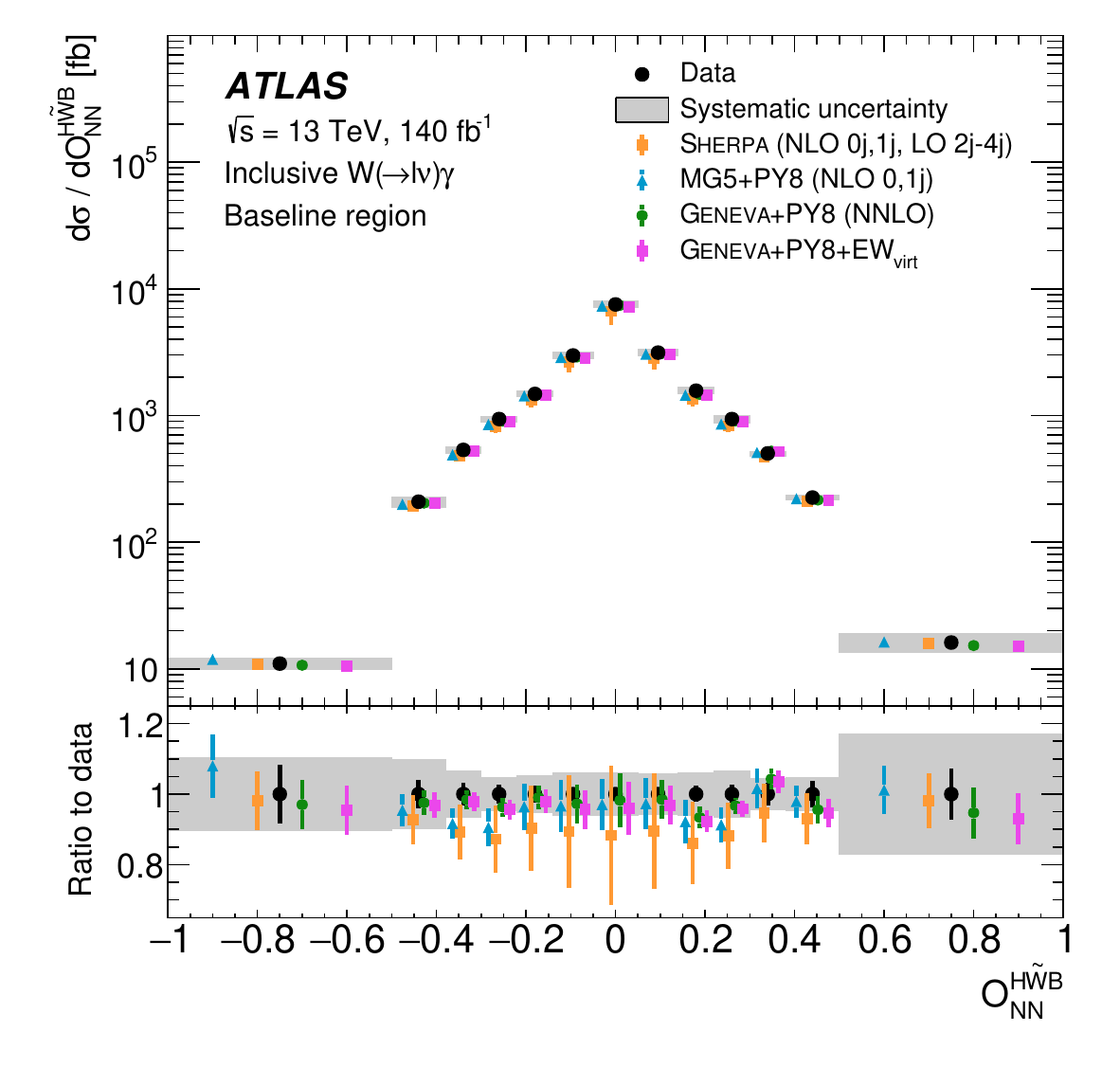}
    \caption{Left: double-differential cross section in the $W$-boson decay angles $\phi_\ell$ and $\theta_\ell$. Right: Differential cross sections as a function of $\mathcal{O}_{NN}{HWB}$ which are CP-sensitive observables constructed from the outputs of neural networks~\cite{ATLAS:Wgamma2026}.}
    \label{fig:ATLAS_Wg}
\end{figure}

The ATLAS Collaboration has published a comprehensive measurement of $W(\to\ell\nu)\gamma$ production at $\sqrt{s}=13\tev$ with $140\fbinv$~\cite{ATLAS:Wgamma2026}. The measurement achieves sufficient precision to discriminate between state-of-the-art theoretical predictions including NNLO QCD (Geneva+PY8) and virtual NLO electroweak corrections (Geneva+PY8+EW$_{\rm virt}$), constituting one of the most stringent tests of higher-order predictions in diboson production to date. Among all the measurements described in the publication I'd like to highlight the following: 

\textbf{Polarisation.} A double-differential cross section in the $W$-boson decay angles $\phi_\ell$ and $\theta_\ell$ is measured, providing direct sensitivity to the $W$ spin-density matrix. The extracted polarisation fractions are in agreement with the SM expectation. See Fig.~\ref{fig:ATLAS_Wg} (left).

\textbf{CP-sensitive observables.} A novel observable \ONN is constructed from the outputs of a multiclass neural network trained to separate the linear interference term from the squared contribution for each CP-odd EFT operator. The observable is defined as $\ONN = P_+ - P_-$, where $P_\pm$ are the probabilities assigned by the network to the two CP-conjugate hypotheses. Applied to the $WW\gamma$ vertex operators $\mathcal{O}_{HWB}$ and $\mathcal{O}_W$, the use of \ONN improves the expected limits on CP-odd dimension-6 operators by a factor of two compared to the use of $\Delta\phi^\gamma_\ell$ alone, while remaining in agreement with the SM. See Fig.~\ref{fig:ATLAS_Wg} (right).

\section{Vector Boson Scattering}
VBS events are selected by requiring two vector bosons in association with two energetic forward jets satisfying large \mjj and \detajj requirements, which suppresses the QCD-induced diboson background and enhances the purely electroweak signal. Following the Run~2 programme, VBS has been experimentally established in all major boson-pair combinations. The results presented here mark the completion of that programme and the beginning of the precision phase, with first Run~3 measurements and dedicated studies of boson polarisation.

\subsection{VBS observation in semileptonic final states (ATLAS)}

The ATLAS Collaboration has reported the observation of electroweak $WW/WZ/ZZ$ production in semileptonic final states using $140\fbinv$ at $\sqrt{s}=13\tev$~\cite{ATLAS:VBSsemi2025}. This is the first observation of VBS in these final states, where one boson decays leptonically and the other hadronically, reconstructed in both resolved and merged topologies. Three lepton channels are combined: 0-lepton ($Z\to\nu\nu$), 1-lepton ($W\to\ell\nu$), and 2-lepton ($Z\to\ell\ell$), denoted $\ell\ell qq$, $\ell\nu qq$, and $\nu\nu qq$ respectively. Signal extraction employs a Recurrent Neural Network (RNN) discriminant. The combined result yields:
\begin{equation}
\muEWK = 1.28\,^{+0.23}_{-0.21}\,,
\end{equation}
with an observed (expected) significance of $7.4\,\sigma$ ($6.1\,\sigma$), constituting a clear observation. A two-dimensional fit in \muEWK versus \muQCD yields $\muQCD = 2.09\,^{+0.60}_{-0.57}$, consistent with the SM within uncertainties, shown in Fig.~\ref{fig:SMP-24-013andATLASsemilep} (left).

The semileptonic topology benefits from an approximately six-fold larger branching fraction than fully leptonic decays and grants access to higher boson transverse momenta, making it complementary to purely leptonic VBS analyses for EFT interpretations.

\subsection{CMS combination of VBS across seven channels}
\begin{figure}
    \centering
    \includegraphics[width=0.47\linewidth]{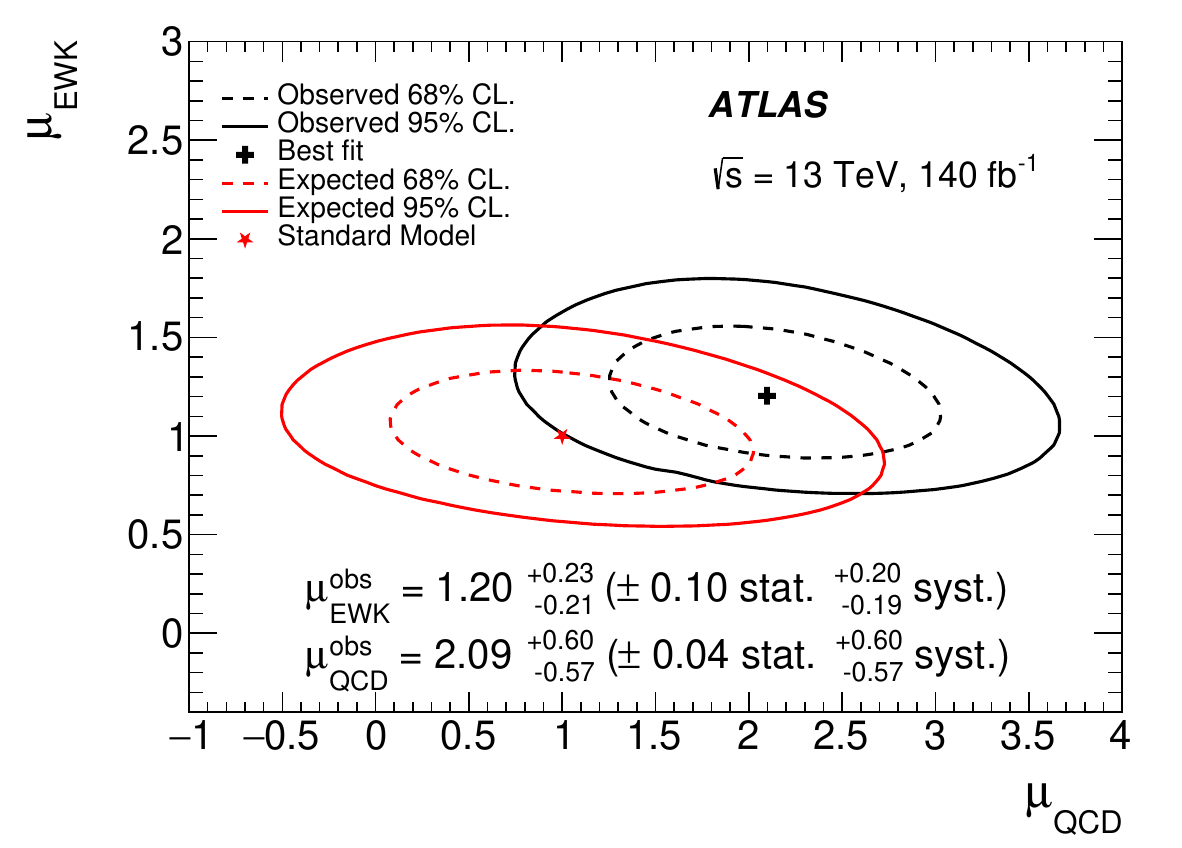}
    \includegraphics[width=0.40\linewidth]{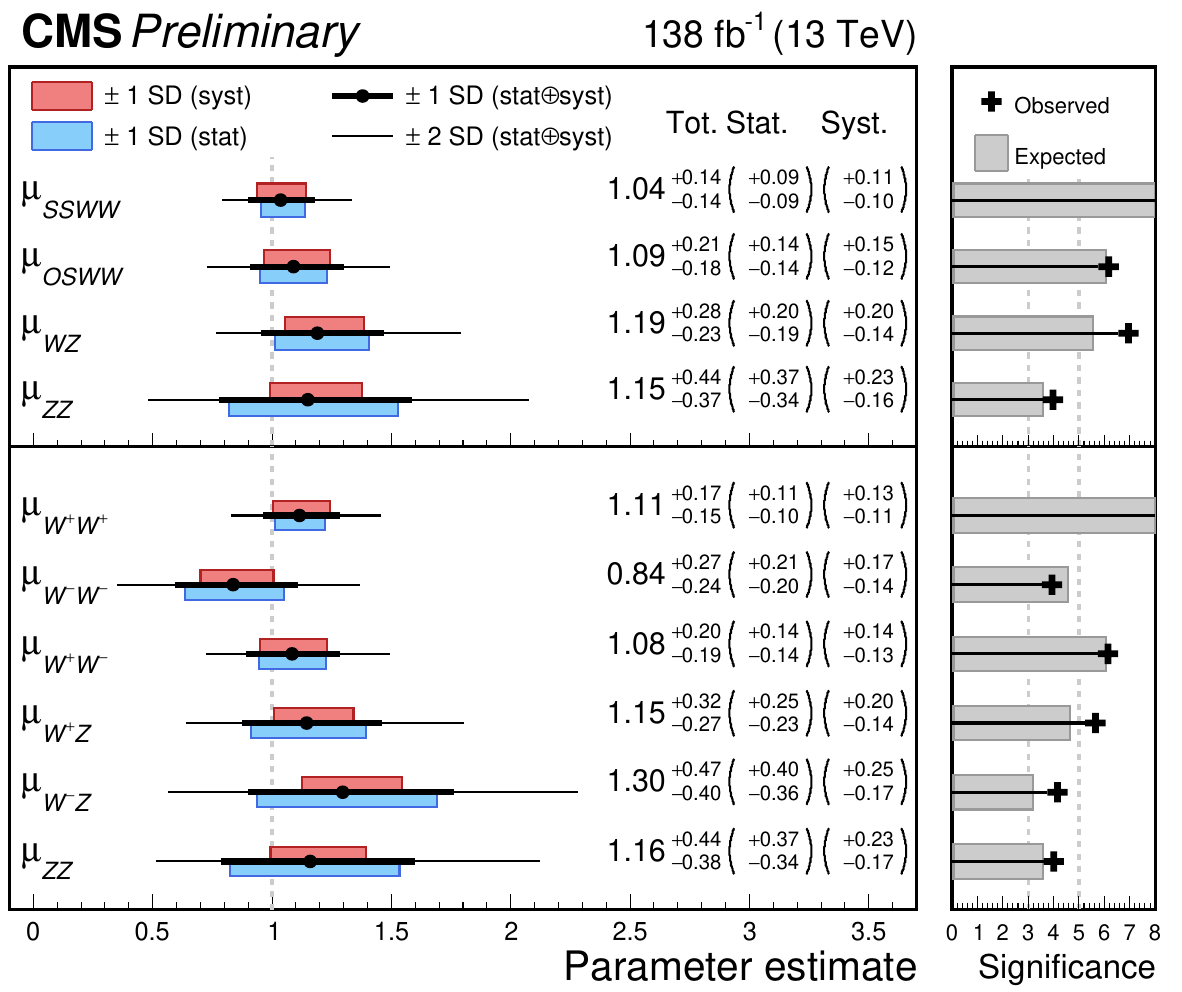} 
    \caption{Left: A two-dimensional fit in \muEWK versus \muQCD yields $\muQCD = 2.09\,^{+0.60}_{-0.57}$, consistent with the SM within uncertainties~\cite{ATLAS:VBSsemi2025}. Right: Measurement of the vector boson scattering (VBS) signal strength modifiers and comparison with Standard Model (SM) predictions. The top (bottom) panel presents results considering four (six) independent signal strengths~\cite{CMS:VBScomb}.}
    \label{fig:SMP-24-013andATLASsemilep}
\end{figure}

The CMS Collaboration has performed a combined measurement of VBS processes across seven channels using $138\fbinv$ at $\sqrt{s}=13\tev$~\cite{CMS:VBScomb}. The analysis performs a simultaneous fit to four fully leptonic channels --- same-sign $W^+W^+$, opposite-sign $WW$ (OSWW), $WZ$, and $ZZ$ --- and three semileptonic channels --- $WV$ and $ZV$. A 4-POI and a 6-POI model measures independent signal strengths for each process. Results are summarized in Figure~\ref{fig:SMP-24-013andATLASsemilep} (right), all consistent with unity. This combination provides a coherent baseline for EFT and aQGC interpretations across the full VBS topology, and constitutes the reference point for future Run~3 updates.

\subsection{Observation of electroweak \texorpdfstring{$ZZ$}{ZZ} production (CMS)}

The final missing piece in the establishment of VBS across all major boson pairs is the observation of electroweak $ZZ$ production. CMS has measured EW $ZZ$ production in the $\ell\ell\nu\nu jj$ final state using $138\fbinv$ at $13\tev$~\cite{CMS:EWZZ}. Signal extraction uses a Graph Neural Network  discriminant. The data/MC agreement after the fit can be seen in Fig.~\ref{fig:CMS_VBS} (left). The measured cross section is:
\begin{equation}
\sigmaEW(ZZjj \to \ell\ell\nu\nu jj) = 0.37\,^{+0.14}_{-0.12}\,(\text{stat.}) \pm 0.06\,(\text{syst.})\fb,
\end{equation}
corresponding to an observed significance of 3.1$\,\sigma$ in the $\ell\ell\nu\nu jj$ channel alone. Combining with the four-lepton $ZZ$ channel yields a combined significance of \textbf{5.0$\,\sigma$}, establishing electroweak $ZZ$ production for the first time. 

\subsection{First VBS measurements at \texorpdfstring{$\sqrt{s}=13.6$}{13.6} TeV (CMS)}

\begin{figure}
    \centering
    \includegraphics[width=0.32\linewidth]{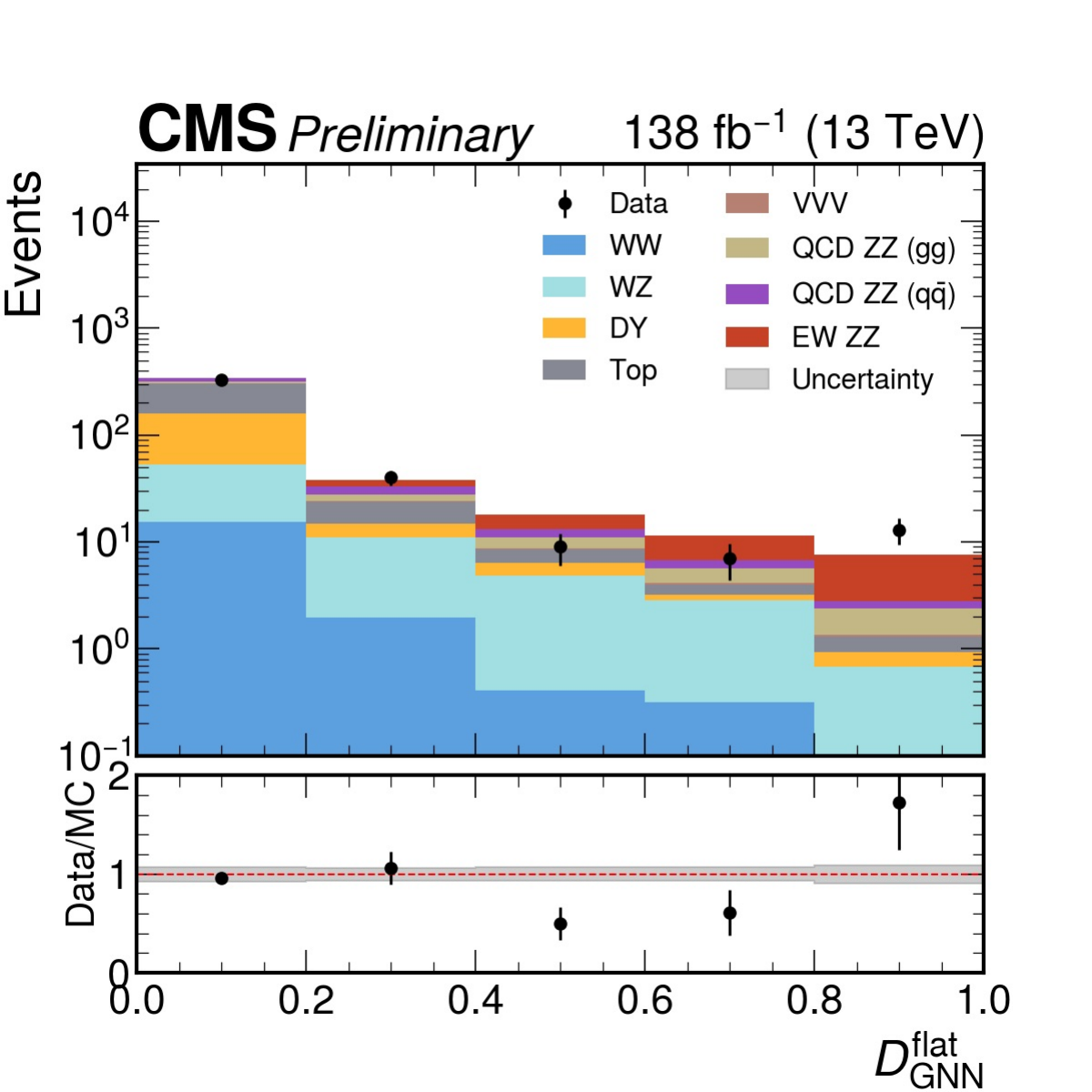} 
    \includegraphics[width=0.32\linewidth]{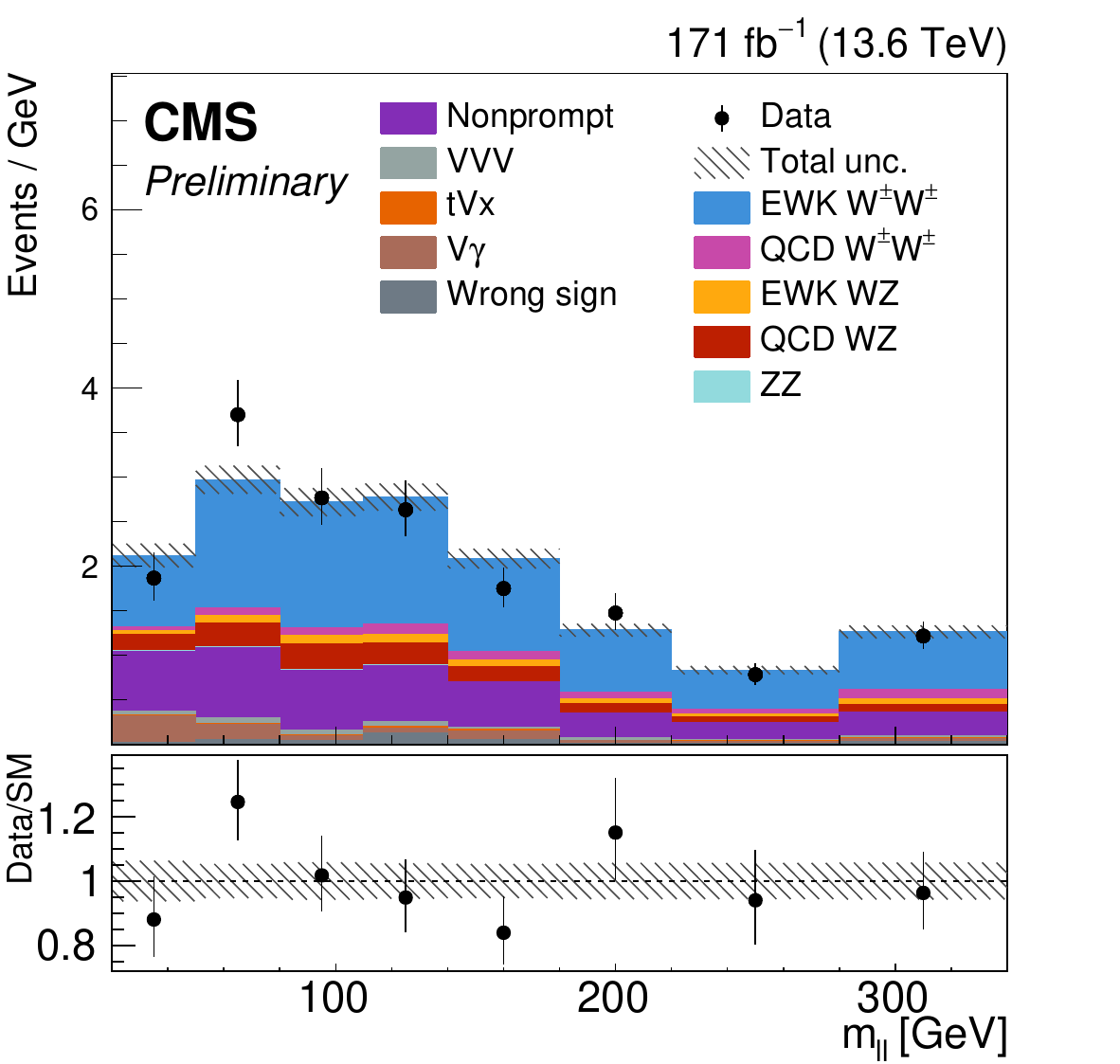}
    \includegraphics[width=0.32\linewidth]{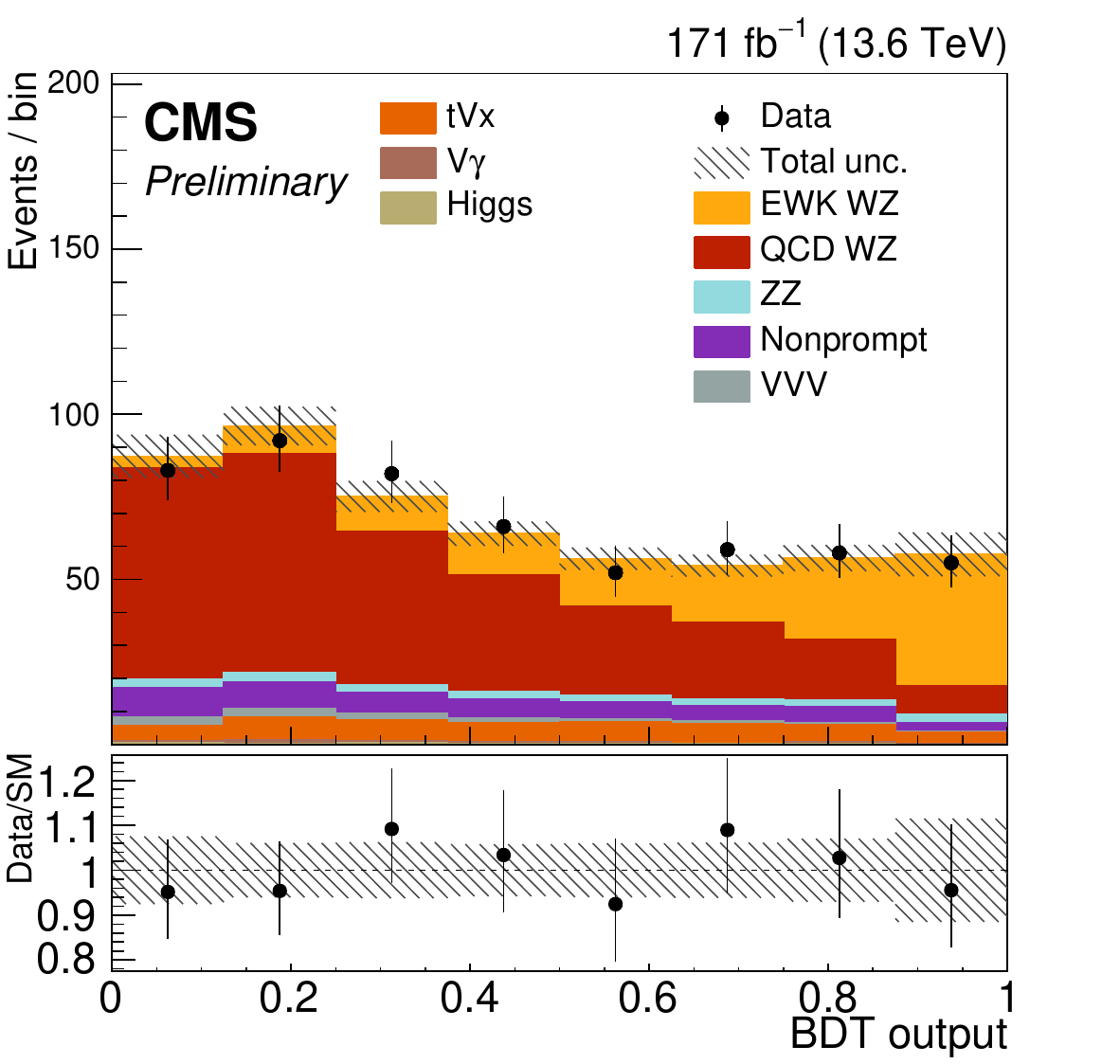}
    \caption{Distribution of the flattened GNN discriminator (left) in the signal region after the maximum-likelihood fit~\cite{CMS:EWZZ}.Distributions of $m_\ell$ in the $W^{\pm}W^{\pm}$ signal region (middle)~\cite{CMS:VBSRun3} and BDT score in the $WZ$ signal region (right)~\cite{CMS:VBSRun3}.}        \label{fig:CMS_VBS}
\end{figure}

The first Run~3 VBS measurements have been performed by CMS in the leptonic $W^{\pm}W^{\pm}$ and $WZ$ channels using $171\fbinv$ at $\sqrt{s}=13.6\tev$~\cite{CMS:VBSRun3}. Both processes are observed with high significance using a BDT-based discriminant for $WZ$ and a simultaneous maximum-likelihood fit ti the $m_\ell$ and $m_{jj}$ distribution. The data/MC agreement after the fit can be seen in Fig.~\ref{fig:CMS_VBS}. The measured inclusive fiducial cross sections are:
\begin{align}
\sigma_{\rm fid}(\text{EW}\,W^{\pm}W^{\pm}jj) &= 3.81 \pm 0.38\fb \quad (>5\,\sigma), \\
\sigma_{\rm fid}(\text{EW}\,WZjj) &= 1.43 \pm 0.26\fb \quad (\approx7\,\sigma),
\end{align}
both in agreement with SM predictions from MadGraph5\_aMC@NLO and Sherpa. Differential distributions in \mjj are also measured for $W^{\pm}W^{\pm}$ production, providing the first probe of the VBS topology at the new centre-of-mass energy. CMS has further shown distributions from the 2022--2025 dataset ($280\fbinv$), which will enable improved sensitivity to aQGC with the full Run~3 statistics.

\section{Triboson Production}

Triboson production provides direct access to quartic gauge couplings through multi-boson radiation diagrams, complementing VBS. The final states are rare and experimentally challenging, requiring simultaneous optimisation across multiple leptonic decay channels. With typical production cross sections of order $10^2$--$10^3\fb$, these processes require the full Run~2 luminosity to achieve observation-level sensitivity. Each observed triboson mode closes a gap in the experimental mapping of the quartic gauge sector: $WWW$ probes the $WWWW$ vertex, $WWZ$ and $VVZ$ probe the $WWZZ$ and $ZZZZ$ structures, and radiative processes such as $WW\gamma$ and $WZ\gamma$ access the $WW\gamma\gamma$ and $WWZ\gamma$ quartic vertices that are forbidden at tree level in the absence of anomalous couplings. The following results represent a remarkable harvest from Run~2, with several processes reaching observation significance simultaneously.

\subsection{Observation of \texorpdfstring{$VVZ$}{VVZ} and evidence for \texorpdfstring{$WWZ$}{WWZ} (ATLAS)}

ATLAS has performed a comprehensive search for $VVZ$ ($V=W,Z$) production using $140\fbinv$ at $13\tev$~\cite{ATLAS:VVZ2025}, analysing both leptonic final states ($WWZ\to\ell\nu\ell\nu\ell\ell$, $WZZ\to\ell\nu\ell\ell\ell\ell$, $ZZZ\to\ell\ell\ell\ell\ell\ell$) and semileptonic final states ($WWZ\to qq\ell\nu\ell\ell$, $WZZ\to\ell\nu qq\ell\ell$). The inclusive $VVZ$ cross section is measured to be:
\begin{equation}
\sigma(pp\to VVZ) = 660\,^{+93}_{-90}\,(\text{stat.})\,^{+88}_{-81}\,(\text{syst.})\fb,
\end{equation}
with an observed (expected) significance of 6.4$\,\sigma$ (4.7$\,\sigma$), representing the first observation of $VVZ$ production. The individual $WWZ$ mode is also extracted, yielding $\sigma(pp\to WWZ) = 442 \pm 94\,(\text{stat.})\,^{+60}_{-52}\,(\text{syst.})\fb$, with an observed (expected) significance of 4.4$\,\sigma$ (3.6$\,\sigma$), constituting evidence for $WWZ$. Both cross sections are consistent with the SM predictions. Limits on dimension-8 Wilson coefficients describing aQGC are also derived.

\subsection{Evidence for \texorpdfstring{$WWZ$}{WWZ} production at 13.6 TeV (CMS)}

CMS has measured the $WWZ$ and $ZH(H\to WW^*)$ production cross sections simultaneously in a novel analysis using $200\fbinv$ combining $\sqrt{s}=13$ and $13.6\tev$ data~\cite{CMS:WWZ2025}. This is the first analysis to separate the nonresonant $WWZ$ and the Higgsstrahlung $ZH$ contributions in a simultaneous fit. The measured signal strengths relative to the SM are:
\begin{equation}
\mu_{WWZ+ZH}(13\tev) = 0.75\,^{+0.34}_{-0.29}, \quad
\mu_{WWZ+ZH}(13.6\tev) = 1.74\,^{+0.71}_{-0.60}.
\end{equation}
The triboson signal at $13.6\tev$ is observed with a significance of 3.8$\,\sigma$ (expected 2.5$\,\sigma$), providing the first evidence for triboson production at this centre-of-mass energy. Combining the two modes and the two center-of-mass energies, the inclusive signal strehnt relative to the SM prediction is measured to be $1.03^{+0.31}_{-0.28}$, with an observed (expected) significance of 4.5 (5.0) standard deviations.

\section{Summary and Outlook}
A rich programme of multiboson and VBS measurements has been reviewed, representing a comprehensive test of the electroweak sector at the LHC. Table~\ref{tab:summary} provides a compact summary of the key results discussed. 

\begin{table}[htb] 
\centering
\small
\renewcommand{\arraystretch}{1.25}
\begin{tabular}{llcc}
\toprule
\textbf{Process} & \textbf{Experiment / Dataset} & \textbf{Sig.} & \textbf{Key result} \\
\midrule
$ZZ\to\ell\ell\nu\nu$ & ATLAS, $140\fbinv$, $13\tev$, \cite{ATLAS:ZZ2025}  & --- & $\sigma_{\rm fid} = 21.0\pm1.0\fb$ \\
$ZZjj\to\ell\ell\nu\nu jj$ & ATLAS, $140\fbinv$, $13\tev$, \cite{ATLAS:ZZ2025}   & --- & $\sigma_{\rm fid} = 0.96\,^{+0.18}_{-0.16}\fb$ \\
$W\gamma$ incl. & ATLAS, $140\fbinv$, $13\tev$, \cite{ATLAS:Wgamma2026}  & --- & NNLO QCD, CP-odd limits $\times$2 \\
EW $VVjj$ (semilept.) & ATLAS, $140\fbinv$, $13\tev$,\cite{ATLAS:VBSsemi2025}  & $7.4\,\sigma$ & $\mu_{\rm EWK} = 1.28^{+0.23}_{-0.21}$ \\
VBS (7-channel comb.) & CMS, $138\fbinv$, $13\tev$,\cite{CMS:VBScomb}  & --- & all $\mu$ consistent with SM \\
EW $ZZjj\to\ell\ell\nu\nu jj$ & CMS, $138\fbinv$, $13\tev$,\cite{CMS:EWZZ}  & $5.0\,\sigma$ & $\sigma_{\rm EW}=0.37^{+0.14}_{-0.12}\pm0.06\fb$ \\
EW $W^\pm W^\pm$ & CMS, $171\fbinv$, $13.6\tev$,\cite{CMS:VBSRun3} & $>5\,\sigma$ & $\sigma_{\rm fid}=3.81\pm0.38\fb$ \\
EW $WZ$ & CMS, $171\fbinv$, $13.6\tev$,\cite{CMS:VBSRun3} & $\approx 7\,\sigma$ & $\sigma_{\rm fid}=1.43\pm0.26\fb$ \\
$VVZ$ & ATLAS, $140\fbinv$, $13\tev$,\cite{ATLAS:VVZ2025}  & $6.4\,\sigma$ & $\sigma=660^{+93}_{-90}{\textrm{(stat.)}}^{+88}_{-81}\textrm{(syst.)}\fb$ \\
$WWZ$ & ATLAS, $140\fbinv$, $13\tev$,\cite{ATLAS:VVZ2025}  & $4.4\,\sigma$ & $\sigma=442\pm94{\textrm{(stat.)}}^{+60}_{-52}\textrm{(syst.)}\fb$ \\
$WWZ$ at $13.6\tev$ & CMS, $200\fbinv$, $13+13.6\tev$,\cite{CMS:WWZ2025} & $3.8\,\sigma$ & $\mu(13.6\tev)=1.74^{+0.71}_{-0.60}$ \\
\bottomrule
\end{tabular}
\caption{Summary of key multiboson and VBS measurements reviewed in this report. Significances quoted are observed. For the EW $ZZjj$ result, the 5.0$\,\sigma$ is from the combination of the $\ell\ell\nu\nu jj$ and $4\ell jj$ channels.}
\label{tab:summary}
\end{table}

Differential $ZZ$ cross sections at ATLAS reach NNLO-level theoretical discrimination and provide new constraints on anomalous neutral couplings. The $W\gamma$ analysis exploits machine-learning-based CP-sensitive observables to directly probe the $W$ spin-density matrix and the $WW\gamma$ vertex, improving expected EFT sensitivity by a factor of two over traditional angular observables.

All major VBS channels are now experimentally established. The ATLAS semileptonic observation ($7.4\,\sigma$) and the CMS establishment of EW $ZZ$ production ($5\,\sigma$ combined) complete the programme. The first Run~3 VBS measurements at $\sqrt{s}=13.6\tev$ mark the transition from discovery to precision cross-section measurement. 

 $VVZ$ has been observed by ATLAS ($6.4\,\sigma$). The first evidence for triboson production at $\sqrt{s}=13.6\tev$ has been reported by CMS in the $WWZ$ channel.

The LHC Run~3 programme, with up to $300\fbinv$ expected at $\sqrt{s}=13.6\tev$, will transform multiboson physics into a genuine precision probe of the electroweak sector. EFT interpretations  will benefit from the increased statistics, particularly in the high-\mjj and high-$p_T$ tails where BSM effects are enhanced. Combined with the High-Luminosity LHC dataset, these measurements will provide among the most stringent indirect constraints on BSM physics at the TeV scale, complementary to direct searches and Higgs coupling measurements.

\section*{References}
\bibliography{moriond}

%%% manually generated bibliography
%\begin{thebibliography}{99}
%\bibitem{ja}C Jarlskog in {\em CP Violation}, ed. C Jarlskog
%(World Scientific, Singapore, 1988).
%\bibitem{ma}L. Maiani, \Journal{\PLB}{62}{183}{1976}.
%\bibitem{bu}J.D. Bjorken and I. Dunietz, \Journal{\PRD}{36}{2109}{1987}.
%\bibitem{bd}C.D. Buchanan {\it et al}, \Journal{\PRD}{45}{4088}{1992}.
%\end{thebibliography}

\end{document}